\documentclass[twocolumn]{article}
\usepackage{times,bm}
\usepackage{amsmath}
\usepackage[affil-it,]{authblk}
\usepackage[pagebackref=false,colorlinks=true,linkcolor=acsblue,citecolor=olive,urlcolor=acsblue]{hyperref}

\usepackage{geometry}
\usepackage[switch*, displaymath,pagewise, mathlines]{lineno}
\usepackage{graphicx}
\usepackage{cite}
\usepackage{amsmath}
\usepackage{amsfonts}
\usepackage{amssymb}
\usepackage{slashed}
\usepackage{subfloat}
\usepackage{slashed}
\usepackage{color}
\usepackage{float}
\usepackage{multirow,multicol}
\usepackage{tabulary}
\usepackage[flushleft]{threeparttable}
\usepackage{pgfplots,subfigure}
\usepackage{xcolor}
\numberwithin{equation}{section}
\usepackage{tikz}
\usepackage{ulem}
\usepackage{hyperref}
\usepackage{nameref}
\usepackage{comment}

\usepgflibrary{arrows}
\usetikzlibrary{shapes.callouts}
\tikzset{
	level/.style   = { thick, },
	connect/.style = { dotted, red   },
	notice/.style  = { draw, rectangle callout, callout relative pointer={#1} },
	label/.style   = { text width=1cm }
}

\definecolor{acsblue}{RGB}{17,76,139}
\definecolor{shadecolor}{RGB}{255,241,204}
\usepackage{letltxmacro}
\makeatletter
\let\oldr@@t\r@@t
\def\r@@t#1#2{%
	\setbox0=\hbox{$\oldr@@t#1{#2\,}$}\dimen0=\ht0
	\advance\dimen0-0.2\ht0
	\setbox2=\hbox{\vrule height\ht0 depth -\dimen0}%
	{\box0\lower0.4pt\box2}}
\LetLtxMacro{\oldsqrt}{\sqrt}
\renewcommand*{\sqrt}[2][\ ]{\oldsqrt[#1]{#2}}
\makeatother
\usepackage{environ}

\begin{document}

\newcommand{{\ri}}{{\rm{i}}}
\newcommand{{\Psibar}}{{\bar{\Psi}}}
\newcommand*\var{\mathit}

\fontsize{8}{9}\selectfont

\title{\mdseries{ Amelino-Camelia DSR effects on charged Dirac oscillators: Modulated spinning magnetic vortices}}

\author{ \textit {\mdseries{Abdullah Guvendi}}$^{\ 1}$\footnote{\textit{ E-mail: abdullah.guvendi@erzurum.edu.tr }}~,~ \textit {\mdseries{Omar Mustafa}}$^{\ 2}$\footnote{\textit{ E-mail: omar.mustafa@emu.edu.tr} }~,~ \textit {\mdseries{Nosratollah Jafari}}$^{\ 3}$$^{\ 4}$$^{\ 5}$\footnote{\textit{ E-mail: nosrat.jafari@fai.kz } }  \\
	\small \textit {$^{\ 1}$\footnotesize Department of Basic Sciences, Erzurum Technical University, 25050, Erzurum, Türkiye}\\
	\small \textit {$^{\ 2}$\footnotesize Department of Physics, Eastern Mediterranean University, 99628, G. Magusa, north Cyprus, Mersin 10 - Türkiye}\\
 \small \textit {$^{\ 3}$\footnotesize Fesenkov Astrophysical Institute, 050020, Almaty, Kazakhstan}\\
 \small \textit {$^{\ 4}$ \footnotesize Al-Farabi Kazakh National University, Al-Farabi av. 71, 050040 Almaty, Kazakhstan}\\
 \small \textit {$^{\ 5}$ \footnotesize Center for Theoretical Physics, Khazar University, 41 Mehseti Street, Baku, AZ1096, Azerbaijan}}
	
\date{}
\maketitle

\begin{abstract} 
This work explores the two-dimensional Dirac oscillator (DO) within the framework of Amelino-Camelia doubly special relativity (DSR), employing a modified Dirac equation that preserves the first-order nature of the relativistic wave equation. By introducing non-minimal couplings, the system provides an exact analytical solution in terms of confluent hypergeometric functions, along with closed-form expressions for the energy spectrum (indulging a Landau-like signature along with accidental spin-degeneracies)-. In the low-energy limit, the results reproduce the well-known two-dimensional Dirac oscillator spectrum, and in the nonrelativistic regime, the results reduce the Schrödinger oscillator spectrum. First-order corrections in this DSR model introduce a mass-splitting term proportional to \(\pm \mathcal{E}_{\circ}/\mathcal{E}_p\), where \(\mathcal{E}_{\circ} = mc^2\) is the rest energy and \(\mathcal{E}_p\) is the Planck energy. These corrections preserve the symmetry between the energies of particles and antiparticles around zero energy, but induce a shift in the energy levels that becomes more significant for higher excited states (\(n > 0\)). By mapping the system to a DSR-deformed charged Dirac oscillator in the presence of an out-of-plane uniform magnetic field, we show that the leading-order Planck-scale corrections vanish at a critical magnetic field \(\mathcal{B}^{c}_{0}\), and as the magnetic field approaches this critical value, the relativistic energy levels approach \(\mathcal{E}_{n,\pm} = \pm \mathcal{E}_{\circ}\). Finally, we identify a previously undetermined feature in two-dimensional charged Dirac oscillator systems in a magnetic field, revealing that the corresponding modes manifest as spinning magnetic vortices.
\end{abstract}

\begin{small}
\begin{center}
\end{center}
\end{small}

%\begin{small}
%PACS numbers: 02.40.-k; 03.65.Ge; 03.65.-w; 04.20.Gz; 04.20.Jb; 04.62.+v;21.45.+v.
%\end{small}
%\bigskip

%\maketitle

\section{ \mdseries{Introduction}}\label{sec:1}

DSR theories arise from modifications of special relativity, incorporating an additional invariant, Planck energy \( E_p = \sqrt{\hbar\, c^5/G} \approx 10^{19} \, \text{GeV} \), alongside the speed of light \( c \). Two prominent models in this area are the Amelino-Camelia DSR \cite{Amelino} and the Magueijo-Smolin (MS) DSR \cite{Magueijo}. One of us (NJ) derived a modified Dirac equation within the Amelino-Camelia DSR framework \cite{NJ}, revealing that this equation can be reduced to Schrödinger-like equations in which particles and antiparticles possess distinct masses. Such frameworks enable exploration of potential DSR effects on well-known quantum systems, offering intriguing insights. Among the most effective tools for examining DSR effects are the exactly solvable relativistic oscillators, including the Dirac oscillator (DO) \cite{Moshinsky}, Klein-Gordon oscillator \cite{o2}, and vector boson oscillator \cite{vbo}.

\vspace{0.10cm}
\setlength{\parindent}{0pt}

The Dirac oscillator (DO) \cite{Moshinsky}, a novel interaction incorporated into the Dirac equation, has established itself as a fundamental framework with significant implications for both theoretical and experimental physics \cite{Franco, guvendi-epjc-1}. This interaction introduces a linear coupling between the momentum and spatial coordinates. In the non-relativistic limit, its solutions reduce to those of the quantum harmonic oscillator, which justifies the term Dirac oscillator. The corresponding electromagnetic potential has been explicitly derived, revealing that the DO describes a system in which the anomalous (chromo)magnetic moment couples with a linearly increasing electric field \cite{Bentez}. This feature makes the DO a compelling model for quark confinement in quantum chromodynamics and for describing the relativistic dynamics of quarks within mesons and baryons \cite{Moshinsky-2}. Its first experimental realization was achieved in a one-dimensional configuration using a tight-binding array of coupled dielectric disks, where the experimental results closely matched theoretical predictions. Since then, the DO has received growing attention across various fields of physics, including mathematical physics \cite{Villalba}, high-energy physics \cite{guvendi-podu}, and quantum optics \cite{Bermudez, Romera}. As an exactly solvable model, the DO provides a valuable theoretical framework for studying the effects of spacetime topology and deformed symmetries on relativistic fermionic systems \cite{Bakke, guvendi-epl, Oliveira}. In particular, it has been extensively investigated within the context of the \( \kappa \)-Poincaré deformation, which generalizes the classical Poincaré algebra into a Hopf algebra structure \cite{Andrade, Chargui, Lukierski}. Although \( \kappa \)-deformed DO has been the subject of several studies, the influence of the Amelino-Camelia formulation of Doubly Special Relativity (DSR) on DO has not yet been addressed, to the best of our knowledge. In this work, we revisit the Dirac equation within the Amelino-Camelia DSR framework and derive the corresponding DSR-modified Dirac oscillator by incorporating leading-order corrections associated with Planck-scale energy. The analysis is further extended to include the presence of an out-of-plane uniform magnetic field.

\vspace{0.10cm}
\setlength{\parindent}{0pt}

The structure of the paper is as follows. In Section \ref{sec:2}, we introduce the DSR-inspired Dirac equation and analyze a two-dimensional Dirac oscillator (DO) through non-minimal coupling. This framework leads to a set of coupled first-order differential equations which are subsequently transformed into a second-order, non-perturbative wave equation. Section \ref{sec:3} presents exact analytical solutions to this equation in terms of confluent hypergeometric functions, resulting in a closed-form expression for the energy spectrum that incorporates leading-order corrections from the Planck energy scale. In Section \ref{sec:4}, we investigate the influence of a uniform out-of-plane magnetic field and its implications for the system's dynamics. Finally, Section \ref{sec:5} summarizes the findings and discusses their physical significance. 

\section{\mdseries{DSR-Modified Dirac Oscillator}}\label{sec:2} 

In this section, we begin by writing the modified Dirac equation within the framework of Amelino-Camelia DSR. We consider a Dirac particle in a flat (2+1)-dimensional spacetime with Cartesian coordinates \((x, y)\), and introduce the corresponding DO equation adapted to this specific DSR setting. The modified Dirac equation takes the form \cite{NJ}:
\begin{equation}
\begin{split}
&\left[\gamma^{t}\partial_t +\gamma^{x}\slashed{\partial}_x\hat{\eta}+\gamma^{y}\slashed{\partial}_y\hat{\eta}+i\tilde{m}\textbf{I}_2\right]\Psi(x^{\mu})=0,\\
&\hat{\eta}=1+\frac{i}{2k}\partial_t.\label{MDE}
\end{split}
\end{equation}
In this context, \(\gamma^{t}\), \(\gamma^{x}\), and \(\gamma^{y}\) denote the generalized Dirac matrices, where \(k = \mathcal{E}_p / \hbar\), and \(\mathcal{E}_p\) represent the Planck energy. The quantity \(\tilde{m} = mc / \hbar\) includes the rest mass \(m\) of the particle and the speed of light \(c\). The spacetime-dependent Dirac spinor \(\Psi(x^{\mu})\) is a two-component spinor, with the index \(\mu = t, x, y\). The DO is described by the following non-minimal substitutions \cite{guvendi-epjc-1}:
\begin{equation}
\slashed{\partial}_x \rightarrow \partial_x + \frac{m\omega_o c}{\hbar} \gamma^{t} x, \quad \slashed{\partial}_y \rightarrow \partial_y + \frac{m\omega_o c}{\hbar} \gamma^{t} y, \label{DOC}
\end{equation}
where \(\omega_o\) denotes the oscillator frequency (or coupling strength). The spacetime is described by the metric \( ds^2 = c^2 dt^2 - dx^2 - dy^2 \) with the signature \((+, -, -)\). In this framework, the generalized Dirac matrices are defined as \(\gamma^t = \frac{1}{c} \sigma_z\), \(\gamma^x = i \sigma_x\), and \(\gamma^y = i \sigma_y\), where the matrices \((\sigma_x, \sigma_y, \sigma_z)\) represent the Pauli spin matrices, consistent with the chosen metric signature \cite{guvendi-epjc-1, guvendi-omar-2024, guvendi-epjc-2}. This formulation enables the decomposition of the two-component Dirac spinor as:
\begin{equation}
\Psi(x^{\mu}) = \tilde{\Psi}(x, y) e^{-i\omega t}, \quad \tilde{\Psi}(x, y) = (\psi_1(x, y), \psi_2(x, y))^{T}, \label{DS}
\end{equation}
where \(\omega\) is the relativistic frequency, and \(^T\) denotes the transpose of the space-dependent spinor. By substituting Eqs. (\ref{DS}) and (\ref{DOC}) in Eq. (\ref{MDE}), we obtain the following set of coupled equations:
\begin{equation}
\begin{split}
&\left[\tilde{\omega} - \tilde{\mu}\right] \psi_1(x, y) - \left[\hat{\partial}_{-} - \kappa \, \hat{r}_{-}\right] \psi_2(x, y) = 0,\\
&\left[\tilde{\omega} + \tilde{\mu}\right] \psi_2(x, y) - \left[\hat{\partial}_{+} + \kappa \, \hat{r}_{+}\right] \psi_1(x, y) = 0, \label{E-set}
\end{split}
\end{equation}
where 
\begin{equation}
\begin{split}
&\hat{\partial}_{\pm} = \partial_{x} \pm i \, \partial_{y}, \quad \hat{r}_{\pm} = x \pm i \, y,\quad \kappa = \frac{m\omega_o}{\hbar},\\
&\tilde{\omega} = \frac{\omega}{c \, \left(1 + \frac{\omega}{2k}\right)},\quad \tilde{\mu} = \frac{mc}{\hbar \, \left(1 + \frac{\omega}{2k}\right)}.
\end{split}
\end{equation}
Here, \(\hat{\partial}_{\pm}\) represent the spin raising (\(+\)) and lowering (\(-\)) operators.

\section{\mdseries{Exact solvability for the DSR-modified DO}}\label{sec:3}

In this section, in alignment with the modified Dirac equation \eqref{MDE}, which incorporates leading-order Planck energy scale corrections, we aim to derive analytical solutions for the modified two-dimensional DO. To achieve this, we will solve the system of equations outlined in Eq. (\ref{E-set}). By exploiting the angular symmetry inherent in polar coordinates, we employ the spin raising and lowering operators as defined in \cite{guvendi-epjc-1}:
\[
\partial_{x} \pm i \partial_{y} = e^{\pm i \theta} \left( \partial_r \pm \frac{i}{r} \partial_{\theta} \right),
\]
alongside the relationships
\[
x \pm i y = e^{\pm i \theta} r.
\]
This transformation allows us to rewrite the set of coupled equations in \eqref{E-set} in the following form
\begin{equation}
\begin{split}
&\left[\tilde{\omega}-\tilde{\mu}\right]\psi_1(r,\theta)-e^{-i\theta}\left[\partial_{r}-\frac{i}{r}\partial_{\theta}-\kappa\,r\right]\psi_2(r,\theta)=0,\\
&\left[\tilde{\omega}+\tilde{\mu}\right]\psi_2(r,\theta)+e^{i\theta}\left[\partial_{r}+\frac{i}{r}\partial_{\theta}+\kappa\,r\right]\psi_1(r,\theta)=0,\label{E-set-T}
\end{split}
\end{equation}
which applies specifically to the transformed spinor components:
\begin{equation*}
\begin{pmatrix}
\psi_1(r,\theta) \\ 
\psi_2(r,\theta)
\end{pmatrix}\Rightarrow
\begin{pmatrix}
e^{i(s-1)\theta}\,\psi_1(r) \\ 
e^{is\theta}\,\,\,\,\,\,\,\,\,\psi_2(r) 
\end{pmatrix},
\end{equation*}
where \( s \) represents the spin. Accordingly, we obtain the following set of coupled equations:
\begin{equation}
\begin{split}
&\left[\tilde{\omega}-\tilde{\mu}\right]\psi_1(r)-\left[\partial_{r}+\frac{s}{r}-\kappa\,r\right]\psi_2(r)=0,\\
&\left[\tilde{\omega}+\tilde{\mu}\right]\psi_2(r)+\left[\partial_{r}+\frac{1-s}{r}+\kappa\,r\right]\psi_1(r)=0.\label{WEE}
\end{split}
\end{equation}
Solving this equation set in favor of \(\psi_2(r)\) leads to the following non-perturbative wave equation:
\begin{equation}
\begin{split}
&\frac{d^2\psi_{2}}{dr^2}+\frac{1}{r}\frac{d\psi_{2}}{dr}+\left[\lambda-\frac{s^2}{r^2}-\kappa^2r^2\right]\psi_{2}=0,\\
&\lambda =\tilde{\omega}^2-\tilde{\mu}^2+2\kappa(s-1).\label{WE}
\end{split}
\end{equation}
This wave equation closely resembles a two-dimensional radial Schrödinger oscillator problem, where \(\kappa\) serves as the oscillator frequency and $s^2/r^2$ identifies a radial repulsive core . By substituting \(\psi_2(r)=\,\frac{1}{\sqrt{r}}\psi(r)\), we further simplify the equation to
\begin{equation}
\frac{d^2\psi(r)}{dr^2}+\left(\lambda -\frac{(s^2-1/4)}{r^2}-\kappa^2 r^2\right)\psi(r)=0. \label{o1}
\end{equation}
This approach enables the investigation of the influence of first-order DSR corrections on the DO and offers a framework for understanding the quantum oscillator behavior in the presence of these modifications. The equation under consideration is the well-known two-dimensional radial Schrödinger oscillator equation, where \( s = \pm |s| = \pm 1/2, \pm 3/2, \dots \) represents the spin (e.g., \cite{oa1, o1, o2}). It admits a regular, finite solution at both \( r = 0 \) and \( r = \infty \) in terms of the confluent hypergeometric function
\begin{equation*}
\begin{split}
&\psi(r) = \mathcal{C} \, r^{|s|+1/2}\, e^{-\frac{\kappa r^2}{2}} \, _1F_1 \left(\frac{2(|s|+1)\kappa - \lambda}{4\kappa},  (|s| + 1),  \kappa r^2 \right)\\
&\Rightarrow \psi_{2}(r)=\mathcal{C} \, r^{|s|}\, e^{-\frac{\kappa r^2}{2}} \, _1F_1 \left(\frac{2(|s|+1)\kappa - \lambda}{4\kappa},  (|s| + 1),  \kappa r^2 \right).\label{o2}
\end{split}
\end{equation*}
In this case, the condition \( \frac{2(|s|+1)\kappa - \lambda}{4\kappa} = -n \) is used to truncate the confluent hypergeometric series to a polynomial of order \( n \geq 0 \) \cite{guvendi-epjc-2,o1}, yielding 
\begin{equation}
\begin{split}
&\tilde{\omega}^2 - \tilde{\mu}^2 = 2\kappa \tilde{n}, \quad \tilde{n} = \left(2n + |s| - s + 1\right),\\
& \hbar^2 \omega^2 - m^2c^4 = 2\hbar^2 c^2 \kappa \tilde{n} \, \left(1 + \frac{\hbar \omega}{2 \hbar k}\right)^2.
\end{split} \label{o3}
\end{equation}
Note that here, \( \hbar \omega = |\mathcal{E}_n| > 0 \), \( \hbar k = \mathcal{E}_p > 0 \), \( \mathcal{E}_\circ = mc^2 \), and \( \left(1 + |\mathcal{E}_n| / 2\mathcal{E}_p\right)^2 = 1 + |\mathcal{E}_n|/\mathcal{E}_p + \mathcal{O}\left(\mathcal{E}_n^2/\mathcal{E}_p^2\right) \). Under these conditions, we obtain
\begin{equation}
    \mathcal{E}_n^2 - \mathcal{E}_\circ^2 = \mathcal{K}_n \left(1 + \frac{|\mathcal{E}_n|}{\mathcal{E}_\circ}\right), \quad \mathcal{K}_n = 2\hbar^2 c^2 \kappa \tilde{n} = 2\mathcal{E}_\circ \mathcal{E}_{\text{osc}} \tilde{n}, \label{o4}
\end{equation}
where $\mathcal{E}_{\text{osc}}=\hbar \omega_\circ$. This leads to a quadratic equation in the form
\begin{equation}
    \mathcal{E}_n^2 - \frac{\mathcal{K}_n}{\mathcal{E}_p}|\mathcal{E}_n| - \left(\mathcal{E}_\circ^2 + \mathcal{K}_n\right) = 0. \label{o5}
\end{equation}
This equation has a delicate nature as \( |\mathcal{E}_n| = \pm \mathcal{E}_{n,\pm} \), where \( \mathcal{E}_{n,+} = +|\mathcal{E}_n| \) and \( \mathcal{E}_{n,-} = -|\mathcal{E}_n| \). Therefore, we obtain
\begin{equation}
     \mathcal{E}_{n,+}^2 - \frac{\mathcal{K}_n}{\mathcal{E}_p}\mathcal{E}_{n,+} - \left(\mathcal{E}_\circ^2 + \mathcal{K}_n\right) = 0, \label{o6}
\end{equation}
for positive energies, and 
\begin{equation}
     \mathcal{E}_{n,-}^2 + \frac{\mathcal{K}_n}{\mathcal{E}_p}\mathcal{E}_{n,-} - \left(\mathcal{E}_\circ^2 + \mathcal{K}_n\right) = 0, \label{o7}
\end{equation}
for negative energies. Thus, we arrive at
\begin{equation}
\begin{split}
\mathcal{E}_{n,\pm} &= \pm \frac{\mathcal{E}_{\text{osc}}\,  \mathcal{E}_\circ}{\mathcal{E}_p} \tilde{n} \pm \mathcal{E}_\circ \sqrt{1 + 2\frac{\mathcal{E}_{\text{osc}}}{\mathcal{E}_\circ} \tilde{n} + \mathcal{O}\left(\mathcal{E}_n^2/\mathcal{E}_p^2\right)}\\ 
&\Rightarrow  \mathcal{E}_{n,\pm} = \pm \frac{\mathcal{E}_{\text{osc}}  \mathcal{E}_\circ}{\mathcal{E}_p} \tilde{n} \pm \mathcal{E}_\circ \sqrt{1 + 2\frac{\mathcal{E}_{\text{osc}}}{\mathcal{E}_\circ} \tilde{n}}.
    \label{o8}
\end{split}
\end{equation}
Here, we observe that \( \tilde{n} = \left(2n + |s| - s + 1\right) \) in (\ref{o8}) suggests accidental degeneracies associated with the spin \( s = \pm |s| \) for each radial quantum number \( n \). Specifically, for \( s = -|s| \), we have \( \tilde{n} = \left(2n + 2|s| + 1\right) \), and for \( s = +|s| \), we have \( \tilde{n} = \left(2n + 1\right) \). This identifies energy level degeneracies so that all \( \mathcal{E}_{n,\pm}^{(s=+|s|)} \)-states converge into the corresponding \( \mathcal{E}_{n,\pm}^{(s=0)} \)-state. Hereby, a Landau-like effect is manifestly observed, in $\tilde{n} = \left(2n + |s| - s + 1\right)$, therefore. In the low-energy (DSR-free) limit (i.e., \( \mathcal{E}_{\circ}/\mathcal{E}_p \rightarrow 0 \)), moreover, the result in (\ref{o8}) simplifies to
\begin{equation}
  \mathcal{E}_{n,\pm} = \pm \mathcal{E}_\circ \sqrt{1 + 2\frac{\mathcal{E}_{\text{osc}}}{\mathcal{E}_\circ} \tilde{n}}. \label{UDO}
\end{equation}
In this regime, the result recovers the well-known energy spectrum for a two-dimensional DO \cite{guvendi-epjc-1,mandal}. However, by accounting for first-order DSR corrections, our finding in (\ref{o8}) reveals that DSR does not induce symmetry breaking between particle and antiparticle energy states around zero energy but instead causes an additional mass-splitting term ($\propto \mathcal{E}_{\circ}/\mathcal{E}_p$). Furthermore, this effect becomes more noticeable at high \( \mathcal{E}_{\text{osc}}\) values, especially for large \( n \). In the limit where \( \mathcal{E}_{\text{osc}} \ll \mathcal{E}_{\circ} \), we can approximate the DO energy expression using a Taylor series expansion for small values of \( \chi = \mathcal{E}_{\text{osc}} / \mathcal{E}_{\circ} \, 2\tilde{n} \). For \( \chi \ll 1 \), we expand the square root as \( \sqrt{1 + \chi} \approx 1 + \chi/2 \). Substituting this approximation into the energy expression yields the well-known energy spectrum \( \tilde{\mathcal{E}}_n \) for a Schrödinger oscillator, given by
\begin{equation}
\tilde{\mathcal{E}}_n \approx \mathcal{E}_{\text{osc}} \tilde{n}, \label{NR_O}
\end{equation}

\section{\mdseries{ Effects of uniform magnetic fields}}\label{sec:4}

In this section, we investigate the influence of a uniform out-of-plane magnetic field \( \mathcal{B}_{0} \), oriented along the \( z \)-axis. For this configuration, the non-zero components of the electromagnetic vector potential \( \mathcal{A}_{\mu} \) are conveniently chosen in the symmetric gauge, as detailed in \cite{guvendi-epjc-1,mandal}:
\[
\mathcal{A}_{x} =-\frac{\mathcal{B}_{0}}{2}\, y,\quad \mathcal{A}_{y} =\frac{\mathcal{B}_{0}}{2}\, x.
\]
With this choice of vector potential, the system of equations in Eq. (\ref{WEE}) becomes:
\begin{equation}
\begin{split}
&\left[\tilde{\omega}-\tilde{\mu}\right] \psi_1(r) - \left[\partial_{r} + \frac{s}{r} - \tilde{\kappa} r\right] \psi_2(r) = 0, \\
&\left[\tilde{\omega} + \tilde{\mu}\right] \psi_2(r) + \left[\partial_{r} + \frac{1-s}{r} + \tilde{\kappa} r\right] \psi_1(r) = 0, \label{E-set-Final}
\end{split}
\end{equation}
where
\[
\tilde{\kappa} = \frac{m \Omega}{\hbar}, \quad \Omega = \left( \omega_{o} - \frac{\omega_c}{2} \right), \quad \omega_c = \frac{|e|\mathcal{B}_{0}}{mc},
\]
with \( \omega_c \) representing the cyclotron frequency and \( e \) the electric charge \cite{guvendi-epjc-1,mandal}. By substituting \( \kappa \) with \( \tilde{\kappa} \), as discussed in the previous section, we derive Eq. (\ref{WEE}). Following the steps outlined above, we determine the particle energies \( \mathcal{E}_{n,+} \) and anti-particle energies \( \mathcal{E}_{n,-} \) for the DSR-modified two-dimensional charged DO within a uniform magnetic field as:
\begin{equation}
\mathcal{E}_{n,\pm}=\pm \frac{\hbar \Omega \mathcal{E}_\circ}{\mathcal{E}_p}\tilde{n}\pm\mathcal{E}_\circ \sqrt{1+2\frac{\hbar\Omega}{\mathcal{E}_\circ}\tilde{n}}. \label{MF}
\end{equation}
This system exhibits several fascinating properties that have profound implications for our understanding of dynamics in (2+1)-dimensional spacetimes. In the absence of the magnetic field, the system is governed by an oscillator frequency \( \omega_o \), which can be seamlessly adapted into a (2+1)-dimensional DO in the presence of a uniform magnetic field. The introduction of the magnetic field modifies the dynamics, resulting in an effective oscillator frequency \( \Omega = \omega_o - \frac{1}{2} \omega_c \), where \( \omega_c \) is the cyclotron frequency. This modification, where the magnetic field reduces the oscillator frequency by half the cyclotron frequency, significantly retains the original dynamics of the system, allowing the preservation of key physical properties \cite{guvendi-epjc-1,mandal}. However, as the magnetic field reaches the critical value \( \mathcal{B}^{c}_{0} = \frac{2\, m\, c\, \omega_o}{|e|} \), the effective frequency \( \Omega \) vanishes, causing a complete cessation of oscillations in the modified DO. This marks a fundamental transition, reducing the modified energy to the total rest mass energy, a highly significant and intriguing outcome. Remarkably, despite the profound influence of the magnetic field, the spectral degeneracies observed in the system remain unaffected.

\begin{figure}[h!]
    \centering
    \includegraphics[scale=0.60]{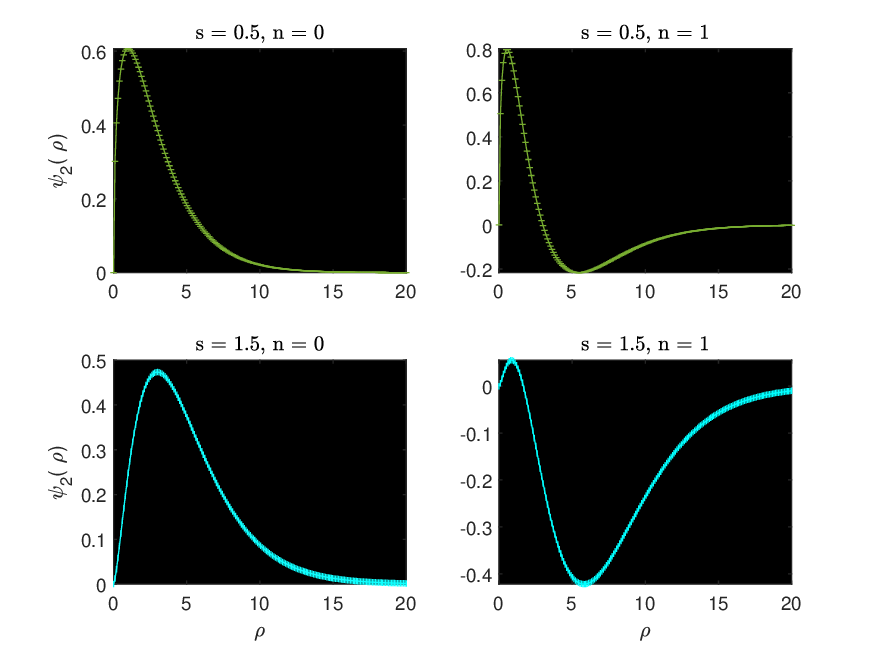}
    \caption{\footnotesize Wave functions \(\psi_{2_{n,s}}(\rho)\) for quantum states \(n = 0\) and \(n = 1\) with spin values \(s = 1/2\) and \(s = 3/2\).}
    \label{fig:a}
\end{figure}

\begin{figure}[h!]
    \centering
    \includegraphics[scale=0.60]{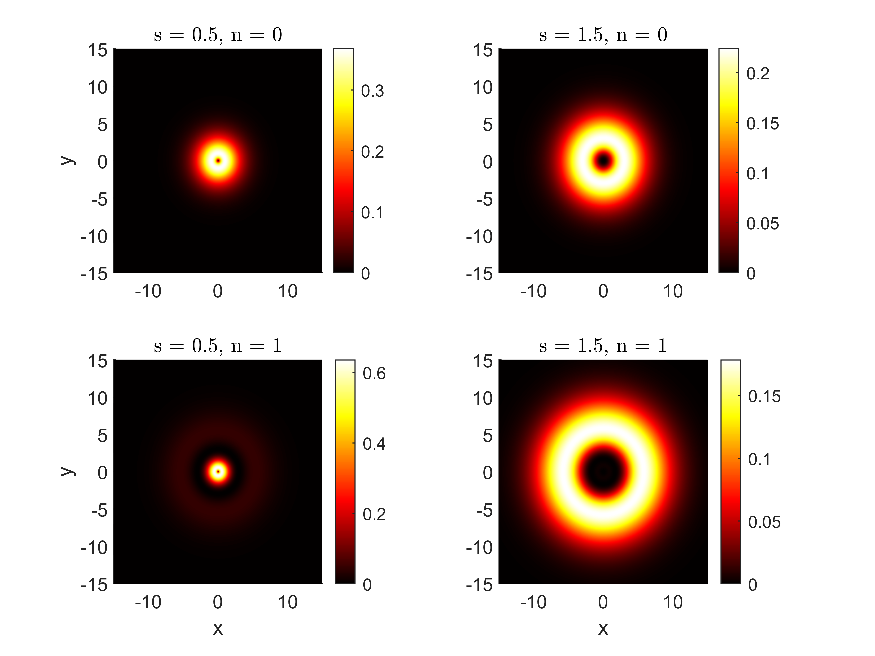}
    \caption{\footnotesize Radial probability density functions \(P_{n,s} = \int |\psi_{2_{n,s}}(\rho)|^2 \, \rho \, d\rho\) for quantum states \(n = 0\) and \(n = 1\), with spin values \(s = 1/2\) and \(s = 3/2\). The probability densities are represented in two-dimensions \((x, y)\), where \(x = \rho \cos(\phi)\) and \(y = \rho \sin(\phi)\).}
    \label{fig:b}
\end{figure}

\section{\mdseries{Summary and discussions}}\label{sec:5}

This study rigorously investigates the dynamics of neutral and charged DO within the Amelino-Camelia DSR in flat (2+1)-dimensional spacetime by introducing a modified DO equation inspired by this specific DSR framework. By incorporating leading-order Planck energy scale corrections, we show that the modified DO equation retains its first-order form, allowing us to derive exact analytical solutions expressed through confluent hypergeometric functions. These solutions reveal first-order DSR corrections in the energy spectra. In the absence of DSR effects (the low-energy limit), our results align with the energy spectra of conventional Dirac and Schrödinger oscillators (in the non-relativistic regime), enabling a precise assessment of DSR’s impact on the energy spectrum. We show that DSR effects induce a mass-splitting effect proportional to the ratio of the rest-mass energy to the Planck energy. This effect becomes more pronounced at higher oscillator frequencies, especially for DO in its excited states. Notably, DSR does not induce symmetry breaking in the particle-antiparticle states near the Dirac point.  

\vspace{0.10cm}
\setlength{\parindent}{0pt}

An intriguing analysis of the system under an out-of-plane uniform magnetic field reveals that the effects of DSR modifications vanish at a critical magnetic field strength, \( B^{c}_0 = \frac{2mc\omega_{\circ}}{|e|} \), at which point the system's energy converges precisely to the rest mass energy. This observation suggests that DSR effects may become fundamentally unobservable when the magnetic field reaches this critical value. The incorporation of DSR modifications into established quantum mechanical models significantly advances our theoretical understanding, particularly in exploring high-energy effects on relativistic quantum systems. The inclusion of a mass-splitting term within this framework offers valuable insight into the impact of DSR on particle-antiparticle states, especially under extreme energy conditions. This approach may bridge the gap between relativistic quantum mechanics and high-energy relativity, where high-energy effects may begin to play a pivotal role. In particular, the emergence of a critical magnetic field strength marks a threshold beyond which the oscillatory behavior of the system is completely suppressed. These findings underscore the robustness and versatility of the proposed model, which provides a firm foundation for further investigations into DSR and its interplay with electromagnetic interactions in relativistic quantum systems. Furthermore, the DO model can be mapped onto well-established quantum optical systems, particularly the Jaynes-Cummings and anti-Jaynes-Cummings models \cite{Bermudez, Romera}, which presents a promising route for experimental verification of the leading-order Planck energy modifications, in principle, within quantum optical frameworks. Additionally, the DSR-modified Dirac equation offers a powerful tool for studying DSR effects in both one-electron atoms and broader many-body systems \cite{guvendi-omar-2024}, further expanding its potential applications in high-energy physics. \\

We would like to emphasize an important, previously undetermined aspect of the dynamics of charged Dirac oscillator systems in a magnetic field. This feature, which is implicitly encoded in the wave function that satisfies the boundary conditions as presented in equation \eqref{o2}, is of considerable significance. In particular, this solution remains valid when the uniform magnetic field is incorporated solely by substituting \(\kappa \rightarrow \tilde{\kappa}\). To proceed, we introduce a new dimensionless variable, \(\rho = \tilde{\kappa} r^2\), where \(\tilde{\kappa}\) has units of inverse length squared. Using this, we can reformulate the physical solution function as:
\[
\psi_{2}(\rho) = \mathcal{C} \, \rho^{|s|} \, e^{-\frac{\rho}{2}} \, _1F_1 \left(-n, (|s| + 1), \rho \right), \quad (n = 0, 1, 2, \dots) \label{Wave-func}
\]
The radial probability density functions are expressed as \( P_{n,s} = \int |\psi_{2_{n,s}}(\rho)|^2 \, \rho \, d\rho \). To illustrate these findings, we present the wave functions and corresponding radial probability densities for the first few quantum states in Figs. \ref{fig:a} and \ref{fig:b}, respectively. Notably, Fig. \ref{fig:b} demonstrates that the modes associated with the charged Dirac oscillator, under the influence of an out-of-plane uniform magnetic field, are exclusively characterized as spinning magnetic vortices (since \(s \neq 0\)). These are manifested as rotating, ring-like modes, a consequence of the factor \(\psi_{2}(\rho) \propto \rho^{|s|}\), even in the low-energy regime. The characteristics observed in these systems cannot be found in one-dimensional DO systems, which only exhibit radial degrees of freedom. Our results indicate that the DSR modification effect, encoded within the quantum number \(n\) through the modified energy spectrum, could, in principle, be observed. However, these effects are quite small as they depend on the ratio \(\mathcal{E}_{\circ}/\mathcal{E}_{p}\).

%\section*{Acknowledgements}
%......

\section*{\small  Data availability}
This manuscript has no associated data.

\section*{\small Conflicts of interest statement}
No conflict of interest declared by the authors.

\section*{\small Acknowledgment}
NJ has been funded by the Science Committee of the Ministry of Science and Higher Education of the Republic of Kazakhstan Program No. BR21881880.

\end{document}